\documentclass[10pt,aps,pre,twocolumn,raggedbottom]{revtex4}

\usepackage{amsfonts}

\usepackage{graphicx}
\def\figspace{\vskip .2 cm }
\def\ts{ {\bf t}(s) }
\def\rs{ {\bf r}(s) }

\def\wr{{\mathcal W}r}
\def\lk{{\mathcal L}k}
\def\tw{{\mathcal T}w}
\def\P{{\mathcal P}}
\def\s{{\mathcal S}}
\def\b{{\mathcal B}}
\def\T{{\mathcal T}}
\def\t{{\bf t}}
\def\C{{\mathcal C}}
\def\ez{\hat {\bf e}_z} 
\def\rh{\hat {\bf u}} 
\def\K{ {\cal K} }

\def\vu{ {\bf u} }
\def\chif{{\chi_F}} 
\def\chifsq{{\chi^2_F}} 
\def\chic{{\chi_C}}
\def\chicsq{{\chi^2_C}}
\def\ca{C\u{a}lug\u{a}reanu}
\def\ess{{\bf \hat e} (s,s')}
\def\tp{{\bf t}_{\perp}  }
\newcommand\moy[2][]{\left\langle{#2}\right\rangle_{#1}}

\begin{document}

\title{Writhing Geometry of Open DNA} \author{V. Rossetto,   A.C.~Maggs  }
\affiliation{
Laboratoire de Physico-Chimie Th\'eorique, UMR CNRS-ESPCI 7083,
  10 rue Vauquelin, F-75231 Paris Cedex 05, France
}
\begin{abstract}  
  Motivated by recent experiments on DNA torsion-force-extension
  characteristics we consider the writhing geometry of open stiff
  molecules.  We exhibit a cyclic motion which allows arbitrarily
  large twisting of the end of a molecule via an activated process.
  This process is suppressed for forces larger than femto-Newtons
  which allows us to show that experiments are sensitive to a
  generalization of the  \ca-White formula for the writhe.  Using
  numerical methods we compare this formulation of the writhe with
  recent analytic calculations.
\end{abstract} 
\maketitle

\section{Introduction}
Recent experiments in which  DNA molecules are manipulated
\cite{ABCS,smith,chatenay} with the help of magnetic beads have led to a
renewed interest in the statistical mechanical properties of the
torsionally stiff wormlike chain
\cite{markosiggia,marko1,nelson,marko,MezardPRL}. In the experiments a
bead is attached to the end of a long DNA molecule; the other end
remains stuck to a surface. The bead is held in a magnetic trap which
allows the simultaneous application of a force, $F$ and couple,
$\Gamma$; one measures the mean distance of the bead from the surface
as a function of $F$ and $\Gamma$.  In an ingenious theoretical paper
it was shown \cite{MezardPRL} that the statistical mechanics of this
problem are related to the problem of a quantum particle in a magnetic
field.  However, a crucial assumption was made in the formulation of
the writhing geometry of the polymer.

The concept of writhe was originally defined in the context of studies
on {\sl closed ribbons} where it forms one part of a topological
invariant, the linking number \cite{cal,white}. In this paper we show
how the writhe can be usefully generalized to study the geometry (and
the mechanical response functions) of open DNA molecules.  This
generalization introduces end corrections to the \ca-White formula for
the writhe. We then perform simulations for the writhe distribution of
an open chain which we compare with the analytic theories.
We show that experiments involving manipulation of DNA with beads in
unconfined geometries have unbounded fluctuations in the measured
torsional angle. Under an exterior torque a bead can rotate an
arbitrarily large angle. When the DNA is under tension this rotation
is an activated process with jumps of $4\pi$ in the mean angle.

Contrary to the calculation performed by Bouchiat and M\'ezard
\cite{MezardPRL} we find that this unbounded response is not removed
by the introduction of an intermediate cut off. However applications
of tensions larger than  femto-Newtons suffice to render the
problem finite in practice \cite{us}. We find that their regularized expressions
{\sl strongly overestimate} the magnitude of writhe fluctuations for
tensions larger than femto-Newtons.

In order to interpret experiments in which the molecule is under
strong tension Moroz and Nelson used a Monge representation to perform
a calculation of writhe fluctuations \cite{nelson}. We numerically
investigate the validity of their results.  We show that a strong
correction to scaling can be expected due to the formation of rare
loops, which give however exceptionally large contributions to the
writhe.  We find that the window of forces in which simple analytic
theories, based on the Monge representation, are valid is rather
small.

In single molecule experiments, self-avoidance confines the polymer to
a single invariant knot. Analytic theories are unable to estimate the
error in writhe due to summing over both knotted and unknotted
configurations.  We investigate this question numerically.

\section{Writhe and Linking Number}
\subsection{Closed curves}
\label{writhe3}
The linking number $\lk$ of a closed ribbon is an integer topological
invariant \cite{cal,white}.  It can be decomposed into two parts the
twist, $\tw$ and the writhe, $\wr$:
\begin{equation}
\lk=\wr+\tw.
\end{equation}
This decomposition is useful because the writhe is a
function of the centerline $\rs$ of the ribbon.
\begin{equation}
  \wr = {1\over 4 \pi} \int ds \int ds' 
 {{\bf r}(s) -{\bf r}(s') \over {|{\bf r}(s)-{\bf r}(s')| ^3}}\cdot 
{d {\bf r} (s)\over ds} \times 
 {d {\bf r} (s')\over ds'}.
\label{doublewrithe}
\end{equation}
The linking number is invariant under deformations of the shape which
do not introduce self intersections. If during a deformation the
centerline $\rs$ crosses itself there is a discontinuity of $2$ in
$\lk$ and thus, as $\tw$ is continuous, a discontinuity of 2 in $\wr$.
Fuller \cite{fuller,fuller2} showed the integral of equation
(\ref{doublewrithe}) could be simplified and introduced the expression
\begin{equation}
\wr^F = {1\over 2 \pi} \int {\ez \cdot( \t \times \dot \t)\over 1
 + \t \cdot \ez} \ d s
\label{fullereq}
\end{equation}
where $\ez$ is the direction at both extremities of the open chain. In
this simplification information is lost so that $\wr$ and $\wr^F$ are
related by the equation
\begin{equation}
\wr-\wr^F =0 \quad {\rm mod}  \, 2
\label{modulo}
\end{equation}
For notational convenience let us introduced the angles, 
$\chic=-2\pi\,\wr$  and $\chif=-2\pi\,\wr^F$.  A much more direct approach to
Fuller's result is possible \cite{acm} by noting that the writhe of a
stiff polymer is closely related to the geometric anholonomies
discovered by Berry \cite{berry} in wave phenomena.

Equation~(\ref{doublewrithe}) has a simple geometric
interpretation\cite{arnold}. The projected path of a chain on a plane
with normal $\vu$ can intersect itself.  Each crossing is assigned a
number, $\pm 1$, according to the handedness of the intersection. The
sum of these numbers~$n(\vu)$ is the writhe number for direction~$\vu$
and the expression, eq.~(\ref{doublewrithe}), is equal to the average
of~$n(\vu)$ over all directions.

\subsection{Bead rotation is given by an extended definition of writhe}
\label{absorbtwist}

The above definition of the writhe, with its emphasis on it being part
of a topological invariant hides, to some degree, the interpretation
of $\chic$ as an angle of rotation in many experimental situations. We
shall now show that an extended definition of the writhe, based on
equation (\ref{doublewrithe}) is the writhe contribution to the bead
rotation that is measured in DNA twisting experiments. Rather similar
arguments have also been given in \cite{marko} for a polymer between
two planes, we give hear an extended derivation to point out the exact
limits of the result.

\begin{figure}[tb]
\figspace
  \includegraphics[scale=.30] {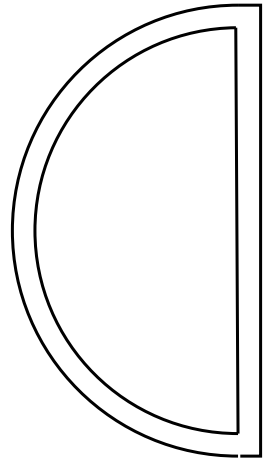} 
\caption{
  A flat ribbon with zero linking number.  }
\label{flat}
\end{figure}

\begin{figure}[tb]
\figspace
  \includegraphics[scale=.3] {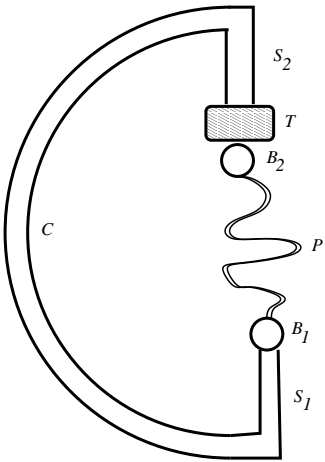} 
\caption{ Deform the ribbon of figure~\ref{flat} as in the diagram: The initial
  and final tangents of the polymer section $\P$ are parallel. Long
  arms $\s_1$ and $\s_2$ attach the polymer to a closing loop $C$. The
  experimentally important reference beads $\b_1$ and $\b_2$ are
  attached to the two ends of the polymer. The twist is absorbed {\em
    exclusively} in the section $\T$.  Note the beads are {\em not} in
  general aligned vertically.  }
\label{deform}
\end{figure}

Consider the planar ribbon in figure~\ref{flat}.  The linking number,
$\lk$, is zero thus
\begin{equation}\wr+\tw=0.
\end{equation} We shall use this closed ribbon as a reference
configuration to calculate the writhe of an {\em open} filament for
which the initial and final tangents are parallel {\em i.e.} for
$\t(0)=\t(L)$, where $L$ is the polymer length. We do this using the
geometry of figure~\ref{deform}.  The polymer $\P$ which is also a
ribbon is embedded in a construction consisting of two long straight
sections $\s_1$, $ \s_2 $ and a closing loop $\C$.  We shall apply
eq.~(\ref{doublewrithe}) to the constructions of
figures~\ref{flat} and~\ref{deform} then take the length of the
straight sections to infinity.

There are beads attached to each end of the polymer, joining on to
the straight sections. These two beads are our experimental reference.
We shall hold the lower bead $\b_1$ stationary and let the upper bead
$\b_2$ rotate due to the writhe of the polymer. The final part of our
construction is the ``twist absorber'' $\T$. This is imagined as being
a joint, or section of the straight ribbon which twists freely to
absorb the writhe generated by the polymer. Let the twist of the
polymer be zero so that {\em all} twist appears in $\T$.

Start from the reference state of figure~\ref{flat} and deform
continuously to an arbitrary state figure~\ref{deform} without
generating a self intersection of the construct. The writhe can be
calculated with the classic double-integral. Clearly since $\wr+\tw=0$
the writhe which is generated goes into twisting the region $\T$ of
the chain. The total twisting angle is just $-2 \pi \wr= \chic$.  Now
let the straight sections go to infinity.  In this limit the
contribution to the integral of the section $\C$ vanishes.  Thus the
{\em experimental} rotation angle is given by the $\chic$ for the
extended construct with the two straight line sections extending to
infinity \footnote{In the light of this construction it is interesting
  to note that the writhe is a conformal invariant for which the
  addition of a single point at infinity is topologically
  ``natural''.}. 

Consider generating an ensemble of chains with some arbitrary
algorithm. The distribution of the configurations of the chains is
independent of the dynamic process creating them if they are subject
to Boltzmann statistics. If we now choose unknotted configurations and
{\sl demand that they be created via a process which preserves the
  linking number} we conclude that we can uniquely determine the
rotation angle from the writhe of the extended construct.  A crucial
part of this argument is the restriction of the construct of
figure~\ref{deform} to the sector $\lk=0$. We discuss the importance
of this restriction in the next section.

\subsection{Choice of linking sector}
In the deformation from figure~\ref{flat} to figure~\ref{deform} we
wish to conserve the linking number $\lk$. To do so we must firstly
forbid self crossing of the polymer configuration. In any experiment
with DNA in the absence of topology changing enzymes 
this restriction in reasonable. However this condition is {\em
  insufficient} to conserve the topology of the entire construction.
We must also forbid the crossing of the real polymer with the
imaginary line from the end of the chain to infinity.  In a recent
numerical paper \cite{marko} this was achieved by grafting the ends of
the polymer to an external surface rather than beads. Experimentally
we would stop this crossing by introducing a steric hindrance: A fine
fiber in the neighborhood of the manipulating bead would work
perfectly. We show below, however, that even this level of steric
hindrance is not needed in the experiments as they are presently
performed.

\begin{figure}[tb]
\figspace
  \includegraphics[scale=.25] {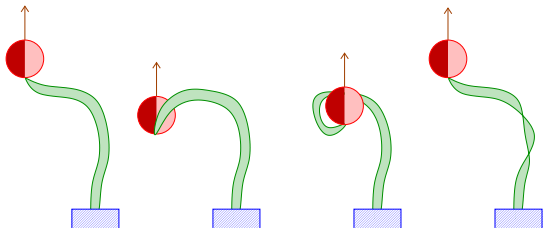} 
\caption{Demonstration of a cyclic motion of DNA leading 
  to introduction of a twist of $4 \pi$.  In the first and fourth
  figure the shape of the molecule is identical. Between the second
  and third figure the vertical extension of the bead passes through
  the molecule leading to a discontinuity in the linking number.  The
  tricky three dimensional geometry is most easily understood by
  repeating this sequence with the help of a belt or ribbon.}
\label{fourpi}
\end{figure}

If the chain is allowed to bend back in such a way that is passes
through the line to infinity there is a discontinuity of $ \Delta\,
\lk = \pm 2$.  This allows one to construct a cycle of motions in
which the shape of the filament undergoes a cycle coming {\em exactly}
back to its original shape while the bead undergoes a rotation of $4
\pi$ about the vertical axis, figure \ref{fourpi}.  This result is the
origin of a common amphitheater demonstration of the importance of
spinor representations of the rotation group: If one holds a plate
horizontally in the palm of one's hand one can spin it about a
vertical axis by performing a suitable contortion of the arm.  Against
all intuition the plate can be turned an arbitrarily large angle.
Photographs of Feynman performing this demonstration are to be found
in \cite{feynman}. Each cycle of the arm again gives rise to a
rotation of $4 \pi$ of the plate.

This unbounded torsional fluctuation is an activated process: When the
DNA is under tension the bead moves against the magnetic force a
distance comparable to its diameter in order to force a loop over the
point of attachment. For micron sized beads this introduces a
characteristic scale of the force of $k_BT/\mu\mathrm{m} \sim
\mathrm{fN}$.  For forces of more than a few femto-Newtons the process
is exponentially rare and suppressed.  The natural time scale for
attempts at crossing the barrier is given by the Zimm time of the bead
$\tau \sim d^3 \eta/k_BT$ with $d$ the bead size, which is seconds.

This force is {\sl extremely} low, the widely studied crossover from
random coil to semiflexible behavior in the force-length
characteristics of DNA occur at a force scale of $k_B T/\ell_p$ with
$\ell_p \approx 53nm$ the persistence length of DNA. This intrinsic
force scale for DNA much higher than that which is needed to conserve
linking number with a $1\mu m$ bead.  We conclude that most
experiments with tense DNA are performed in a regime where a high
energy barrier leads to conservation of the linking number of the
construct of figure 2.  However, there is {\em always} a small
probability of passage by this barrier so that under torsion the true
steady state of a DNA molecule is a state of cyclic motion in which
the bead rotates by a series of activated jumps.

Because of these considerations we assume that all topological
invariants are conserved during experiments, including linking number
and the knot configuration of the extended construct of figure 2.

\subsection{Analytic expressions for the extended writhe}

From the above discussion we are lead to the calculation of the writhe
of the extended construct, figure 2. The double integral,
eq.~(\ref{doublewrithe}), on the interior of the chain is identical to
that commonly used for closed DNA, we denote the corresponding angle
by $\chi_{int}$.

The integral from one point on the linear extension to infinity plus
the second point on the polymer can be simplified as follows: The
integral is evaluated by noticing that it is the spherical
area, swept out by the vector
\begin{equation}
\ess =  {   {   {\bf r}(s) -{\bf r}(s')} 
\over {|{\bf r}(s)-{\bf r}(s')|}} 
\end{equation} 
as $s$ and $s'$ vary.  Consider, now, a chain, $\rs$. Place the origin
at ${\bf r}(0)$.  Let us now place $s$ in the interior of the polymer
and $s'$ on the extension of the polymer to infinity which is directed
in the direction $\ez$.  Consider, now, $s'=0$. The curve $\rh
(s)={\bf \hat e}(s,0)$ is a spherical curve. As we now let $s'$ vary
from $0$ to infinity we sweep out the area between $\rh$ and the point
$\ez$. This spherical area can be written as
\begin{equation}
\chi_{end} = \ez\cdot\int\,ds\,{\rh \times  d\rh/ds \over { 1+ \rh\cdot\ez}}.
\label{endwritheeq}
\end{equation}
then $\chi_C = \chi_{end} + \chi_{int}$.
A similar result has been found \cite{MezardPRL} by direct integration
of eq.~(\ref{doublewrithe}).  We are rather remarkably back to a
problem in statistical mechanics which is very close that of the
Fuller formulation, eq.~(\ref{fullereq}), of the writhe fluctuations.
We convert eq.~(\ref{endwritheeq}) to spherical coordinates and
find:
\begin{equation}
\chi_{end} = \int \,ds\, (1-\cos{\theta}) {d \phi \over {ds}} 
\label{endwritheeq2}
\end{equation}
The singularity at $1+ \rh\cdot\ez =0$ corresponds to the limit
$\cos{\theta}=-1$.

\subsection{Twisting}
Experiments are sensitive to the sum of the writhing and twisting
fluctuations in a polymer. If we now allow the excitation of the
torsional modes in the polymer of figure~\ref{deform} the total
rotation between the two ends is the sum of the writhing angle $\chic$
and rotation due to internally excited twisting motions.  In order to
compare the twisting fluctuations to writhing fluctuations, we
define~$\ell_t=C/k_BT$ the torsional length, where~$C$ is the
torsional modulus \cite{MezardPRL}.  This twisting mode is unmodified
by the writhing geometry, at least in the simplest models of DNA
elasticity. The mean square twisting angle, $\chi_T$ is given by
\begin{math}
\langle \chi_T^2 \rangle = {L \over \ell_t}
\end{math}

\section{Artefacts in the Fuller formulation of Writhe}

Recent analytic work \cite{MezardPRL,nelson} has been based on the
simplified formulation of eq.~(\ref{fullereq}). It is much easier
to treat analytically than the full double integral in
equation~(\ref{doublewrithe}) due to a direct mapping onto quantum
mechanics: The bending energy of a stiff beam in the slender body
approximation of elasticity is given by
\begin{equation}
E = {A\over 2} \int \left( {d \t (s)\over ds}  \right)^2 \ ds
\label{stiff} \ ,
\end{equation}
where $A$ is the bending modulus linked to the persistence length by
$\ell_p=A/k_B T$

To calculate the partition function one must now sum over all paths
\begin{equation}
{\mathcal Z} = \sum_{paths} e^{-E/k_B T}
\label{pathint}
\end{equation}
The sum for the partition function is clearly closely related to path
integrals studied in quantum mechanics.  Formally the energy in
eq.~(\ref{stiff}) looks like the kinetic energy of a free
particle moving on a sphere. From the sum of paths in equation
(\ref{pathint}) one derives a Fokker-Planck equation which is entirely
analogous to the Schroedinger equation for a particle on a sphere:
\begin{equation}
{\partial P(\t ,s) \over \partial s}
=
{1\over 2 \ell_p}
\nabla^2P(\t ,s) \ ,
\label{fp}
\end{equation}
where $\nabla^2$ is the Laplacian operator on the sphere.  Here
$P(\t,s)$ is the probability of finding the chain oriented in the
direction $\t$ at the point $s$.  As a function of $s$ the vector
${\bf t}$ ``diffuses'' with diffusion coefficient which varies as
$1/\ell_p$.

Let us now consider the writhe calculated with Fuller's formula for
such ``diffusing'' paths.  There is a singularity in equation
(\ref{fullereq}) near $\t= -\ez$.  In the neighborhood of this
direction the integral measures twice the winding of the random walk
about the point $-\ez$. The winding of a random walk has
singular properties in the two dimensional plane and on a sphere
\cite{orsay}.  In particular there is logarithmic divergence in the
winding properties in the continuum limit.  It is this winding number
divergence that was picked up in the analytic calculation of
\cite{MezardPRL} and necessitated an intermediate scale cut off in the
calculation.

One of the principle conclusions of this paper is that these winding
number singularities are not present in the distribution of the writhe
determined experimentally: As shown above in the case of constrained
linking number it is the \ca-White expression for the writhe which
gives the exact twisting angle. The Fuller expression differs by an
arbitrary factor of $4 n \pi$ due to the winding number singularities
not present in the original formulation.
We shall demonstrate numerically that it is the passage
from~(\ref{doublewrithe}) to~(\ref{fullereq}) which introduces these
singularities.  Use of~(\ref{fullereq}) will be shown to lead to a
substantial error in the calculation of~$\chic$ with an overestimate
by a factor of~$2.5$ for a discretization corresponding to DNA.

\section{Numerical Methods}

Given the difficulty of treating the full \ca-White expression for the
writhe analytically we decided to proceed by numerical exploration of
the distributions of writhe implied by the \ca-White and Fuller
formalisms. In particular we look for the Cauchy tail predicted
analytically \cite{MezardPRL} in the writhe distribution function.
The existence of such a tail implies the absence of a reasonable
continuum limit of the wormlike chain and continuous evolution of the
response functions as a function of a microscopic cut off. We shall
conclude that the \ca-White formulation remains finite even in the
continuum limit.

In our numerical investigations we shall be particularly interested in
the fluctuations in the writhe of the open chain, as a function of the
tension.  Due to the algorithm used in generating the chains we are
unable to generate chains in an ensemble with an imposed torsional
couple. Our results are thus always for $\Gamma=0$.  The torsional
fluctuation are, as usual, related to the linear response of the chain
via the fluctuation-dissipation theorem.

\subsection{Numerical Calculation of Writhe}

A number of methods are available for the calculation of the writhe of
a discretized polymer \cite{linkwrithe}.  We calculate the discretized
versions of the integrals of eq.~(\ref{doublewrithe}) and equation
(\ref{fullereq}) by recognizing that they are both areas on a unit
sphere. In the Fuller formulation it is the area enclosed by the curve
$\ts$. For the discretize chain the tangent curve becomes a series of
link directions, $\t_i$, corresponding to points on a sphere. One 
connects these points by geodesics and then sum over the area of the
triangles formed by two successive tangent vectors and $\ez$.
\begin{equation}
\chif=\sum_n {\cal A} (\ez, \t_n ,\t_{n+1})
\end{equation}
where $\cal A$ is the area of the spherical triangle defined by the
three vectors.

We have already noted that the full expression,
eq.~(\ref{doublewrithe}) corresponds to the area swept out by $\hat
{\bf e}(s,s')$. For two links forming a discrete chain this defines a
spherical rectangle. We calculate its area by decomposing it into two
spherical triangles.

In both calculations we calculate the area of a spherical triangle by
using l'Huilier's expression
\begin{eqnarray}
{\cal A} = 4 \arctan(&(&\tan((a+b+c)/4)\times\\ \nonumber
                  & & \tan((c-a+b)/4)\times \\ \nonumber
                   & &\tan((c+a-b)/4)\times\\ 
                  & & \tan((a+b-c)/4))^{1/2}) \nonumber
\end{eqnarray}
where, $a$, $b$ and $c$ are the lengths of the sides of a spherical
triangle.  For our purposes the triangle has to be oriented, the area
can be either positive or negative.

A useful cross check in the programming is that for any polymer the
modulo relation of eq.~(\ref{modulo}) must be satisfied despite
very different intermediate results in the calculation. Numerically we
found that the equality held to within $10^{-10}$, when working in
double precision when the extended \ca-White definition of writhe was
compared with the Fuller formulation.

Since the double integral of equation (\ref{doublewrithe}) is reduced
to a double summation this step takes  $O(N^2)$ operations for a
chain of $N$ links. For the long chains studied in our simulations it
is by far the slowest step in the calculation.

\subsection{Link Sector Choice}

The generation of long unknotted chains is numerically difficult; In
our simulations we used an ensemble of chains with only the constraint
$\lk=0$ which, as shown above, is the minimum constraint needed for
unbounded torsional response. Our result can then be directly compared
with existing analytic theories which do not impose any topological
constraints.

The ensemble of $\lk=0$ chains may contain knots.  While
this ensemble may appear physically ``unreasonable'' one must not
forget that knots are rare in the chains that we shall study.  It has
been noted \cite{grosberg} that for a flexible chain unknotted
configurations dominate the statistics of chains even several hundred
Kuhn lengths long.  For the chain lengths that we work with in this
paper the contamination coming from such knotted configurations should
be weak.  At the end of the paper we present a partial investigation
of the influence of knots. We find that they do not modify our
conclusions as to the nature of the continuum limit for writhing
chains. The errors due to the use of a knotted ensemble are
much smaller than the differences between the Fuller and \ca-White
formulation of the writhe.

\subsection{Chain generation}

In order to use the expressions for the writhe given above we are
interested in chains in which the initial and final tangents are
parallel (though the writhe for non parallel configurations does have
a simple generalization \cite{acm}) but for which there is no
constraint on the final position of the chain.  Rather than using a
conventional Monte-Carlo algorithm to generate chains we used a simple
``growth algorithm''.

\subsubsection{Zero force}
We wish to grow chains of length $L$, persistence length $\ell_p$
using a series of links of length $b$.  In the absence of tension we
generate chains by starting from a single link in the $\hat{ \bf e}_z$
direction at the origin.  We then successively add links to the chain
with small random angle increment $\alpha_0 \sim \sqrt{b/\ell_p}$ to
produce a single realization of an equilibrated semiflexible chain.
It is almost certain that this chain does not satisfy the boundary
conditions on the tangent thus we continue growing until the final
tangent is parallel to the initial tangent to within an angle small
compared with $\alpha_0$. We keep the polymer in our ensemble if the
length of the polymer is less than $1.05\times L$, otherwise the whole
configuration is rejected and the process restarted from the first
link. The configurations are then used to calculate the writhing
distributions.  The curves that we generate are somewhat ``imperfect''
since they are due to a mixture of lengths.  This admixture of chain
lengths plays no role, however, in our analysis of the asymptotic
distribution of writhe.  There is no self avoidance in this code; it
can be shown from a Flory argument that self avoidance is a weak
effect in semiflexible chains of moderate length.

\subsubsection{Finite force}
In the presence of an external force the algorithm is slightly more
complicated. We proceed by noting that the partition function of a
chain under tension can be expressed in a very similar manner to the
partition function of a flexible polymer in an external potential
\cite{doiEdwards}.  We proceed by simulating the equation
\begin{equation}
{\partial Z \over \partial s}
=
{1\over 2 \ell_p}
\nabla^2Z + f \cos (\theta)\, Z\ ,
\label{fp2}
\end{equation}
where $\theta$ is the angle between the direction of the force and the
local tangent to the polymer, and~$f=F/k_BT$. $Z( s, \theta)$
corresponds the number of configurations in which the chain points in
the direction $\theta$ after a distance $s$.

As proposed in \cite{orland} we introduce a {\em pool} of several
chains which we grow simultaneously. As each link is added there is
angular diffusion as described above and a second process of birth or
death of chains in the pool in order to account for the force. If $f
\cos (\theta)$ is positive then it is considered to be a growth rate
for reproduction of chains in the pool. If $f \cos (\theta)$ is
negative the chain is stochastically destroyed with the appropriate
probability. We also manage the total pool size as in \cite{orland}.
At the end of a pool growth, we destroy the chains that do not satisfy
the condition for the tangent vector to be parallel to
within~$\alpha_0$ at both extremities.

There are several sources of error possible with the algorithm.  The
most difficult to evaluate is the effect of the finite pool size.  The
result of a single run is an ensemble of several configurations
together with a total weight coming from the management of the pool
size. For sufficiently large pool sizes this weight is the same for
each realization of the growth process. For small pool sizes, however,
this weight undergoes important fluctuations. To calculate a
correlation function from an ensemble of pools we chose to select a
single chain from each pool and performed a simple average over at
least $10,000$ pools. Since we had no, a priory method of estimating
errors from this procedure we experimented with the pool size for
several different values of the force. We found that even when varying
the pools size from as low as $20$ chains to $2000$ chains the
estimates of the mean square writhe were stable within a few per cent.
In our production runs we chose a value of $50$ chains per pool.  An
alternative procedure would weight each pool according to the true
variation of the total weight of each simulation.

A systematic difference between a discretized chain and a continuous
curve also occurs. Our principle aim is to understand experiments on
DNA\cite{ABCS,smith} therefore we choose $\ell_p=53\,\mathrm{nm}$.  We
chose for $b$ the half pitch of a single helix, $1.8\,\mathrm{nm}$,
yielding $\ell_p\simeq30b$. This is also comparable to the diameter of
the molecule.  In order to study the convergence of the writhe to the
continuum limit we shall also perform some simulations with $\ell_p/b\ 
\gg 30$ (see section~\ref{comment}).

\section{Asymptotics of Writhe in open flexible chains}
Before presenting our numerical results on semiflexible chains we wish
to explore the scaling behavior of the writhe of a polymer described
as a freely jointed chain with $N$ links.  This allows us to
understand the length scales and the structures which are important in
determining the writhe of a long molecule.  Many of these results are
already known for {\em closed} chains however we wish to demonstrate
that the end corrections in eq.~(\ref{endwritheeq}) are subdominant in
the limit of very long chains.

\subsection{Scaling arguments for flexible chains}
We shall start with the interpretation of the writhe as a signed area
swept out by the vector $\ess$: Consider two links~$i$ and~$j$ of
length $a$ separated by a distance $R_{ij}$.  This area scales as
$A_{ij} \sim \pm (a/R_{ij})^2$ when $R_{ij}\gg a$.  When $a\ll R_{ij}$
the area is bounded above by $2\pi$.  Clearly when one averages over
random walks the integral in equation (\ref{doublewrithe}) gives zero.
The mean squared writhe can be estimated as
\begin{equation}
\langle \chicsq \rangle \sim \sum_{pairs} A_{ij}^2 P(R_{ij})
\end{equation}
where $P(R_{ij})$ is the pair distribution function for the polymer.
For a three dimensional Gaussian polymer this function scales as
\begin{equation}
P(R_{ij}) \sim 1/R_{ij}
\end{equation}
for lengths smaller than the radius of gyration, $R_G \sim \sqrt{aL}$
of the polymer.  Approximating the sum by an integral we find the
internal contribution to the writhe,
\begin{equation}
 \moy {\chi^2_{int}} \sim {L \over a^4} \int^{R_G}_a {a^4 \over R^4} {a \over R}
    \quad R^2\, dR \sim {L\over a} \label{estimate}
\end{equation}
This integral converges at large distances, but is divergent at small
distances.  We conclude that the writhe integral is dominated by the
cut off scale $a$. For a semiflexible polymer the cut--off $a$
corresponds to the persistence length $\ell_p$. It is thus the
structure at this length scale which dominates the writhe of the
molecule. The average crossing number is defined in a very similar
manner to the writhe except the unsigned area $|A_{ij}|$ is summed
over rather than the signed area.  This has a non-zero mean and scales
in the following way in a Gaussian polymer
\begin{equation}
 \moy{\chi^X} \sim {L\over a^4} \int {a^2 \over R^2} {a \over R} 
  \quad R^2\, dR \sim {L \over a} \log{L \over a}.
  \label{wrL}
\end{equation}
leading to a logarithmic divergence.  Thus in the crossing properties
of an arbitrary projection of the polymer we expect all length scales
are important.

Finally, eq.~(\ref{endwritheeq}), the end correction to the writhe is
a sum of random areas $A_i\sim \pm a/R_i$ where $R_i$ is now the
distance between the end of the polymer and the single link $i$. The
corresponding estimate of the mean square writhe is thus
 \begin{equation}
 \moy{{\chi^2_{end}}}
  \sim  {1\over a^3}\int {a^2 \over R^2} {a \over R} \quad R^2\, dR
  \sim \log{L\over a}.
\label{endscale}
\end{equation}
giving a logarithmic contribution with the structure of the whole
molecule being important.

When self avoidance is introduced in the problem we use the result
that $P(R) \sim 1/R^{4/3}$\cite{degennes} to show that the integral in
equation (\ref{estimate}) is still dominated by short length scales.
The results for the end correction and average crossing number are
however modified. They too become sensitive to structure at short
wavelengths in the polymer.  Thus from eq.~(\ref{estimate}) and
eq.~(\ref{endscale}) the end corrections remain small for long chains
and are subdominant,
\begin{math}
  \langle \chi^2_{int} \rangle \gg \langle \chi^2_{end} \rangle
\end{math}.

Given the importance of this end correction in the interpretation of
the experiments we now present a more rigorous study of the end
correction and confirm its subdominant nature compared with the
internal contributions to the writhe.

\subsection{Magnitude of End Corrections}

In order to further study the scaling behavior of the end corrections
we  examine the limit $L/\ell_p$ large. We thus study the problem
of a freely jointed chain rather than the semiflexible chain and
disprove arguments of \cite{reply} that the dominant singularities in
the writhe of a semiflexible chain come from ends due to the formal
analogies between eq.~(\ref{fullereq}) and eq.~(\ref{endwritheeq}).
Numerical results (not shown here) on semiflexible chains lead to the
same conclusions.

Let us now make a hypothesis that the asymptotics of $\chi_{end}$ are
dominated by the largest contributions to the integrand of
eq.~(\ref{endwritheeq2}) then
\begin{equation}
\chi_{end} \sim \int \,ds\, {d \phi \over {ds}}
\end{equation}
which is the winding number of the polymer about $\ez$.  The winding
number of a semiflexible random walk, $W$, about an infinite line
scales as
\begin{equation}
W^2 \sim \log^2{(L/\ell_p)}
\end{equation}
where $L$ is the length of chain and $\ell_p$ the persistence length.
The Cauchy singularity appearing in this problem is regularized by the
stiffness of the chain, $\ell_p$.

\begin{figure}[tb]
\figspace
  \includegraphics[scale=.45] {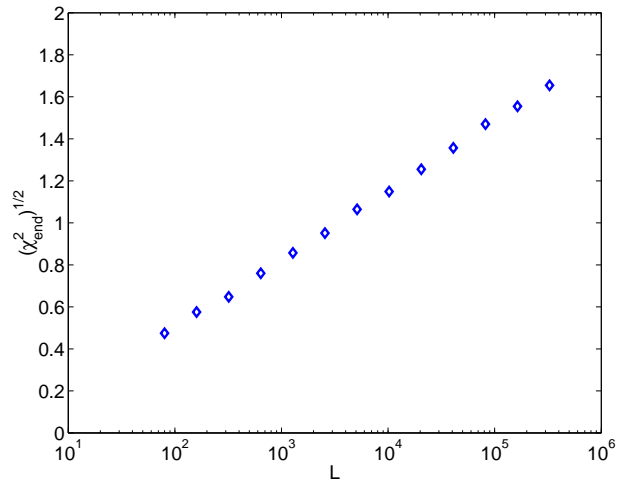} 
  \caption{
    Variation of end correction, $\sqrt{\langle {\chi^2_{end}\rangle
        }}$ , with $L$ showing scaling identical to that of the
    winding of a random walk about an infinite line.  Freely jointed
    chains with up to 320,000 links.  }
\label{endwrithe}
\end{figure}

To check this hypothesis (and to improve on the very rough scaling
analysis given above, eq.~(\ref{endscale})) we generated,
figure~\ref{endwrithe}, a large number of random walks for a freely
jointed chain.  We then plot $\sqrt{\langle {\chi^2_{end}\rangle }}$
as a function of $\log{L}$ and look for a straight line.
We conclude that the end corrections to the writhe are comparable to
\begin{math}
  \moy{\chi^2_{end}} = \log^2{(L/\ell_p)}
\end{math}. 
They are negligible compared with the internal contributions which
vary as
\begin{math}
  \moy{\chi^2_{int}} \sim (L/\ell_p)
\end{math}
for $L \gg \ell_p$.

Despite the similarity between eq.~(\ref{endwritheeq}) and
eq.~(\ref{fullereq}) we find very different scaling for
\begin{math}
  \moy{\chifsq} \sim L \log{(\ell_p/b)}
\end{math} 
and
\begin{math}
  \moy{\chi_{end}^2} \sim \log^2 {(L/\ell_p)}
\end{math}.
At first sight this might seem rather surprising, however in the
Fuller formulation one averages over realizations of two dimensional
random walks in the surface of the sphere whereas in equation
(\ref{endwritheeq}) we average over {\em three} dimensional random
walks {\em projected} onto a sphere.  The statistical weights are
different even if the functions are similar.

\section{Writhe distribution of semiflexible polymers}
\subsection{Short Molecules}
For short filaments of length $L \ll \ell_p$ the distribution of
writhe calculated with the extended \ca-White formula and the Fuller
formula are indistinguishable; ambiguities due to winding about the
pole are exponentially rare.  The writhe distribution of a open
polymer with parallel tangents at each end in the limit $L/\ell_p \ll
1$ is given by the L\'evy \cite{levy,acm} formula for the distribution
of the area enclosed by random walk in a plane
\begin{equation}
{ P} (\chic) =  { \ell_p \over {2 L} } {1 \over \cosh^2(\chic\,  \ell_p/L )}
\quad .
\label{writhedis}
\end{equation}
By generating an ensemble of $10^6$ short chains and binning the
writhe we verified that our code was able to reproduce this
result.

\subsection{Long molecules, zero tension}
\begin{figure}[tb]
  \figspace
  \includegraphics[scale=.54] {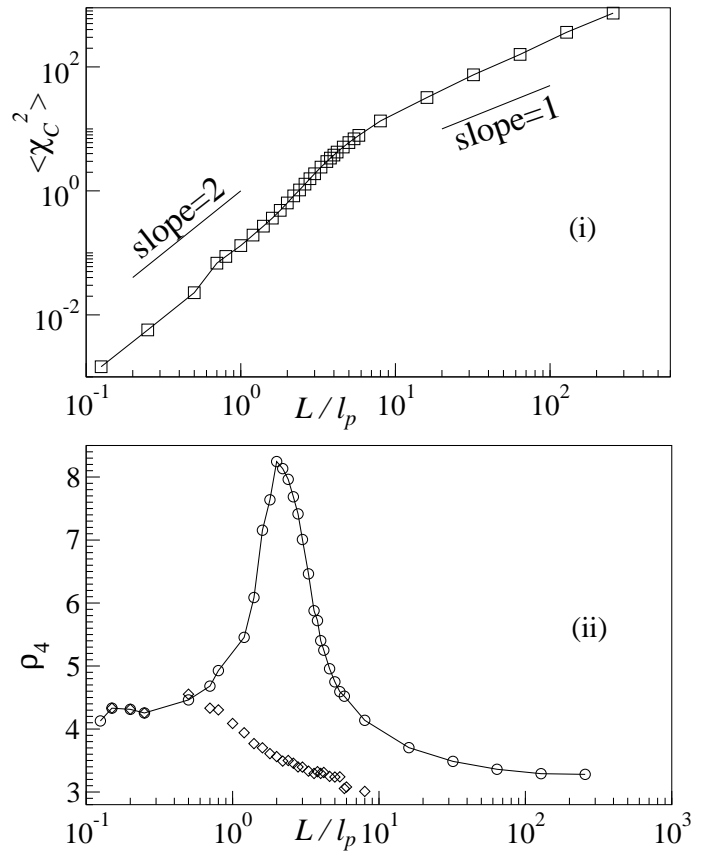} 
   \caption{
     Top: $\moy{\chicsq}$ as a function of~$L$.  Bottom: $\rho_4$ as a
     function of $L/\ell_p$.  When $L/\ell_p\to0$ we find
     $\rho_4\simeq21/5$.  There is a peak around~$L/\ell_p\simeq2$,
     due to the presence of a small number of loops (typically one or
     two), which enlarge the probability distribution of the writhe.
     For larger values of $L/\ell_p$ $\rho_4 \rightarrow 3$
     corresponding to a Gaussian distribution.  $\Diamond$ corresponds
     to chains with $n_S=0$.  100,000 chains for each point,
     $\ell_p=1000b$ for short chains and $\ell_p=50b$ for long chains.
     }
   \label{fig:carac}
\end{figure}

We have characterized the evolution of the writhe properties of a
chain as a function of its length, $L$. We plot, figure
\ref{fig:carac} top, $\moy{\chicsq}$ as a function of $L$ for $\ell_p$
fixed.  For small $L$, eq.~(\ref{writhedis}), we have
$\moy{\chicsq}\sim L^2$ and for long chains, we have
$\moy{\chicsq}\sim L$, eq.~(\ref{wrL}).

As a second characteristic of the writhe distribution, figure
\ref{fig:carac} bottom, we consider $\rho_4= \moy{ (x - \moy{x} )^4 } / \moy
{ (x - \moy{x} )^2 }^2 $ related to the kurtosis, calculated for a
probability distribution~$p(x)$.  If $p$ is Gaussian, then~$\rho_4=3$.
For the distribution eq.~(\ref{writhedis}) $\rho_4=21/5$.  We see
numerically that $\rho_4=21/5$ for small~$L$ and tends slowly to~$3$.
However, there is a large peak around~$L/\ell_p=2$. The very strong
non-monotonic behavior in figure \ref{fig:carac} bottom is most
striking; we shall now explain the origin of this feature.

Let's call {\em back-facing segments\/} the sections of the chain
along which $\t\cdot\ez<0$. We define, $n_S$, the number of such
segments for a chain.  If we compute $\rho_4$ with only configurations
for which $n_S=0$ the peak disappears.  We understand that the peak is
due to the formation of loops in the chain which can, sometimes,
completely dominate the writhing properties of a chain. Such important
feature in the distribution are clearly missed in any Monge
description of a chain.

\label{def:nS}
From these simulations, we also estimated the coefficient of
proportionality~$K$ between~$\moy\chicsq$ and $L/\ell_p$ for very long
chains when the distribution of writhe has converged to a Gaussian:
\begin{equation}
\moy\chicsq\simeq K\frac{L}{\ell_p}\qquad K=2.85\pm0.03
\label{alpha}
\end{equation}
the coefficient is similar in magnitude to that found
\cite{vologo,klenin} for closed chains, even though our model for the
chain is somewhat different.

\subsection{Convergence of  the writhe distribution}
\begin{figure}[tb]
\figspace
  \includegraphics[scale=.5]{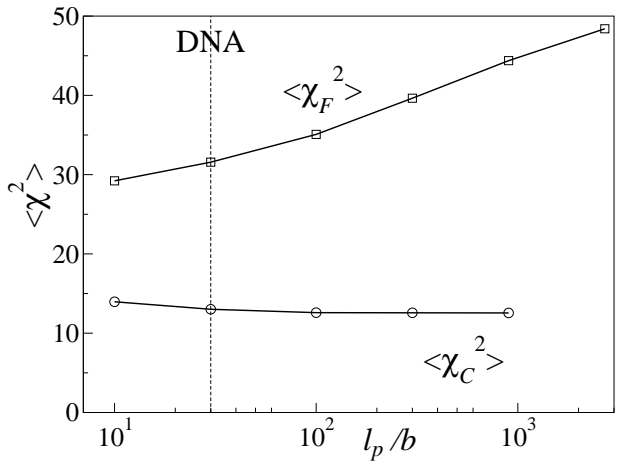} 
\caption{
  Variance of the writhe computed with two different formulations as a
  function of discretization. $\chif$ evolves logarithmicly, whilst
  $\chic$ converges to a constant.  DNA corresponds to $\ell_p/b\simeq
  30$, for which $\moy\chifsq/\moy\chicsq =2.4$.  $L/\ell_p=8$, 200000
  chains per point.  The statistical error is smaller than symbols'
  size.}
\label{modulofig}
\end{figure}

In order to characterize the asymptotic distribution of the writhe,
and study the importance of the Cauchy tail in Fuller formulation, we
have performed simulations on a series of chains of length $L=
8\ell_p$.  With chains of this length knots remain rather rare whilst
the energetic barrier needed for a chain oriented in the direction
$\ez$ to wind about the direction $-\ez$ is only a few $k_BT$. We are
thus sensitive to the winding singularities of the Fuller formulation.
In our simulations we vary the discretization so that there are $L/\ell_p=$
10, 30, 100, 300, 900, 2700 links per persistence length and we
generate 200000 independent configurations for each value of $L/ell_p$.  We
plot the variance of $\chif$ and $\chic$ as a function of the
discretization in figure~\ref{modulofig}.  We observe a continuous
evolution showing a logarithmic divergence for the Fuller formulated
writhe, and a convergence to a stable value for the \ca-White form for
quite moderate values of $L/\ell_p$.
\label{comment}
The ratio between~$\moy\chifsq$ and~$\moy\chicsq$ for DNA value
of~$\ell_p=30b$ is 2.4. We also see that the difference between a
chain with 30 links per persistence length with the continuum limit is
small (about 3\%).

A divergence of the angular fluctuations implies the breakdown of
linear response in the continuum limit. Since, however, $\langle \chicsq \rangle$
converges to a finite value we conclude that a microscopic cutoff is
not needed to render a torsionally stiff chain finite.

\subsection{Long tense molecules}
\label{long}
Until now we have ignored the effect of tension on the configuration
of the DNA, except to remark that even very low tensions justify the
use of a conserved linking number in the interpretation of the
experiments.  In this section we indicate how the writhe of DNA varies
in the presence of external forces.

When a molecule is under high tension the molecule is largely aligned
parallel to the external force.  We can use a simplified, quadratic
form for the Hamiltonian \cite{nelson}
\begin{equation}
E ={1\over 2} \int \left\{ \K \left(\partial \tp \right)^2 +
 \Gamma (\tp \wedge \dot \tp \cdot \ez)^2 +k_B T f \tp^2
\right\}
 \quad {\rm d} s
\label{Enels}
\end{equation}
where $\tp$ describes the transverse fluctuations in the direction of
the molecule. We can find the mean squared writhing angle using the
usual methods of equilibrium statistical mechanics.
\begin{equation}
\moy\chifsq = {\partial^2 \log {\cal Z} \over \partial \ (\beta \Gamma)^2} 
\label{nelsonmonge}
\end{equation}
A short calculation gives
\begin{equation}
{\moy\chifsq} = {1 \over 4}\,\sqrt{ 1 \over f \ell_p}
  \quad \frac {L}{\ell_p}.
\label{nelsonlaw}
\end{equation}
This expression can only be expected to be valid for forces such that
$f \ell_p > 1$.

To estimate the writhe of a molecule at low forces we return to the
remark above that the internal contribution to the writhe is dominated
by structure occurring at the scale $\ell_p$. Under low tensile forces
the structure of a polymer is unchanged out to a length scale $\ell_f
= 1/f$. We conclude that under low tension, when $\ell_f \gg \ell_p$
the writhe of a molecule becomes independent of its degree of
elongation and thus the tension, $f$. It is only under the highest
forces when the semiflexible nature of the molecule is sampled that we
see an evolution of the writhe with force.

\begin{figure}[tb]
\figspace
  \includegraphics[scale=.6] {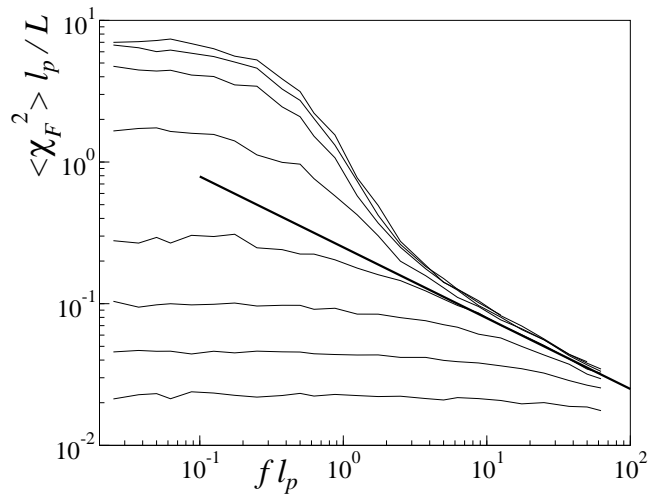} 
\caption{
  $\moy\chifsq \ell_p/L$ as a function of the scaled tension. At large
  forces the result converges to eq.  (\ref{nelsonlaw}). At low
  tensions the writhe is independent of the force. Simulations for
  $\ell_p=250b$.  Curves for $L/\ell_p=$ 0.25, 0.5, 1, 2, 4, 8, 16, 32
  (bottom to top).  The straight line is eq.  (\ref{nelsonlaw}).
\label{lp250}
}
\end{figure}

To validate our code for generating tense molecules we performed a
series of simulations with a chain of persistence length
$\ell_p=250b$, results are shown on figure~\ref{lp250} using the
Fuller expression for the writhe.  At high forces we see that the
curves converge towards the law eq.  (\ref{nelsonlaw}) and that at low
forces the curves saturate as expected from the above scaling
argument. Somewhat surprising however is the rapid crossover which
occurs for forces comparable to $f \sim 1/\ell_p$ when $L/\ell_p \gg
1$.  In the neighborhood of this force there is a pronounced
``shoulder'' on the curve. Convergence to the long chain limit is
rather slow. It is not until $L \sim 60 \ell_p$ that we see a
saturation in writhing curves.  This slow convergence can be
understood rather easily by noting that any chain within $\ell_p$ of
the surface of the polymer coil is in a region of lower than average
density.

\begin{figure}[tb]
  \includegraphics[scale=.6] {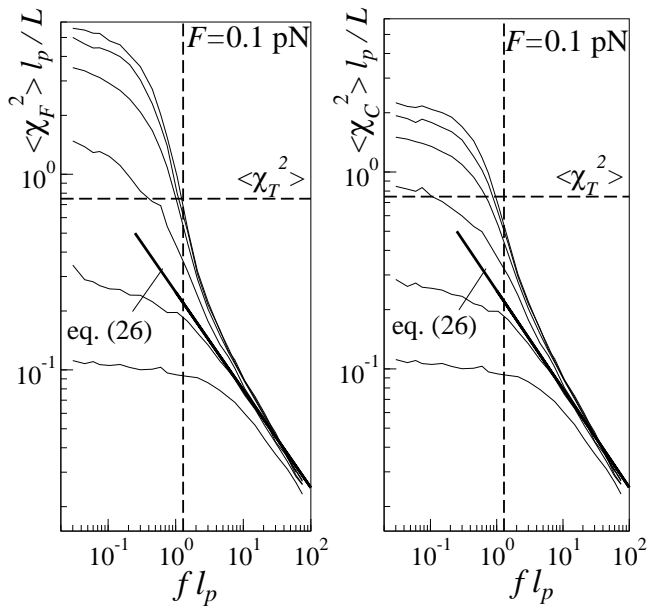} 
\caption{
  $\moy\chifsq \ell_p/L$ (left) and $\moy\chicsq \ell_p/L$ (right) as
  a function of the scaled tension. For longer chains the expected
  regime in $1/\sqrt{f}$ is hidden by a large shoulder from the
  crossover to the low force regime.  Simulations for $\ell_p=30b$.
  Curves for $L/\ell_p=$ 1, 2, 4, 8, 16 and 32 (bottom to top). The
  straight bold line is equation~(\ref{nelsonlaw}).  At zero force,
  the ratio~$\moy\chifsq/\moy\chicsq$ tends to~2.5 as~$L/\ell_p$
  grows.  The horizontal dashed line is the amplitude of fluctuation
  due to twist fluctuations when $\ell_t/\ell_p=1.5$.  The vertical
  dashed line corresponds to~$F=0.1\;\mathrm{pN}$ for DNA.
\label{lp30} }
\end{figure}

We have performed a series of simulations on different levels of
discretization of the polymers. We find that as the discretization
becomes coarser the large shoulder dominates over the law in
$1/\sqrt{f}$ for the mean square writhe.  This is illustrated in
figure~\ref{lp30} where we use 30 links per persistence length for the
discrete chain.  The chain of length $L=\ell_p$ still displays a
regime in agreement with eq. (\ref{nelsonlaw}). However with longer
chains the regime in $1/\sqrt{f}$ is overwhelmed by the crossover to
the low tension regime. Eq. (\ref{nelsonlaw}) substantially
underestimates the writhe fluctuations in the domain out to $f
\ell_p=10$ corresponding to very large forces of $\sim1 \, pN$. At
such forces other corrections come into play, including the chiral
nature of the DNA chain.

It is interesting to note that the Fuller formulation gives results
which are useful over a larger window of forces.  The Fuller and
\ca-White curves are very similar down to tensions $f\ell_p \sim2$;
analytic calculations based on the Fuller formulation can be expected
to be useful for tensions larger than $0.2\, pN$.  At very low forces
the fluctuations are rather different.  The ratio
$\moy\chifsq/\moy\chicsq$ is about 2.5; use of the Fuller formulation
strongly over estimates the torsional response functions.

\subsection{Origin of the shoulder}

\begin{figure}[tb]
\figspace
\begin{center}
\includegraphics[scale=0.5]{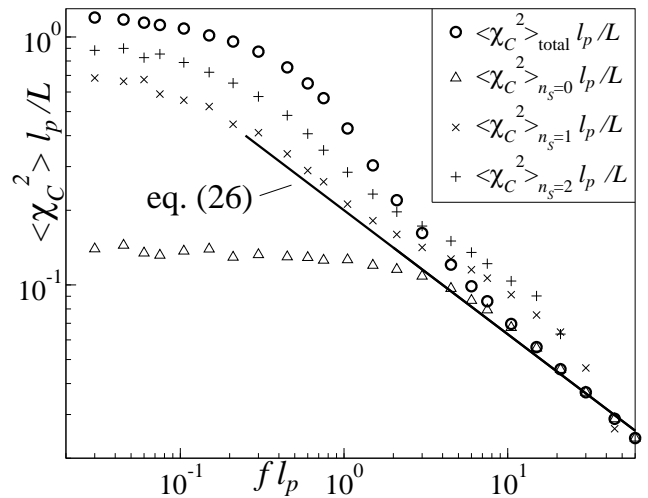} 
\caption{$\moy{\chicsq}\ell_p/L$ as a function of $f\ell_p$.
  Circles all chains.  Triangles, averaged from no back-facing
  segment.  We have plotted the curves for $n_S=1$, $\times$ and
  $n_S=2$, ~($+$).  Each point computed from 300,000 configurations of
  length $8 \ell_p$ with $\ell_p=30b$.  Equation~(\ref{nelsonlaw})
  solid line }
\label{signe}
\end{center}
\end{figure}

To understand the origin of the shoulder which appears in figure 6 we
have classified the different configurations as a function of the
number of back-facing segments  and computed
$\moy{\chicsq}$ for subsets with $n_S$ fixed.  The results are shown in
figure~\ref{signe} for $\ell_p/b=30$.  When $f\ell_p\sim 1$ we see
that the curves for different values of~$n_S$ separate. $n_S=0$ curves
do not exhibit a shoulder.

The origin of the shoulder in figure~\ref{lp30} is essentially the
same as the origin of the peak in figure~\ref{fig:carac}.  Under a
large force, the configurations with back-facing segments are
statistically rare because they have a large energy cost.  Their
contribution to $\moy{\chicsq}$ is, however, important.

\section{Writhe and Knots}
\label{noeuds}
Clearly, self avoidance of the chain implies that topological
invariants are constrained \cite{reply}, including $\lk$.  Another
topological invariant that should be constrained is the knot
configuration. The algorithm that we have used until now leads to an
ensemble containing both knotted and unknotted chains, even if
$\lk=0$. We now present a preliminary investigation on the influence
of knots on the distribution of writhe.  This question is related to
the problem of the closure of the chain: If the polymer is allowed to
pass through the line to infinity, a knot may appear, change or
disappear. We thus continue with the assumption that the tension of
the DNA is sufficiently high that the extended construct of figure 2
remains in the same state throughout an experiment.

\begin{figure}[tb]
  \figspace
\begin{center}
\includegraphics[scale=0.6]{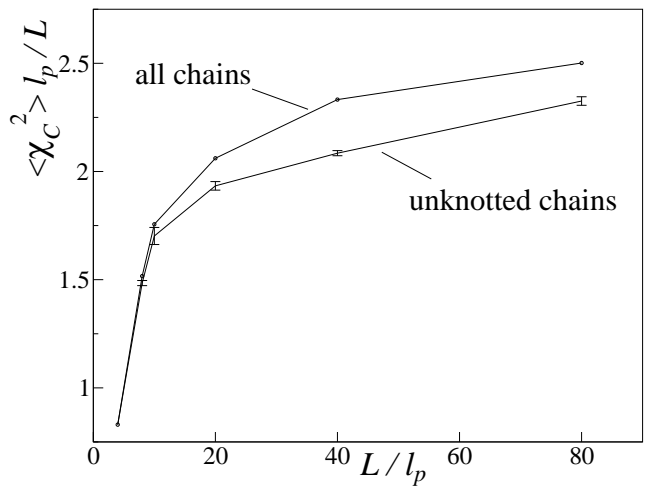} 
\caption{Fluctuations of~$\chic$ for all chains, compared
  with unknotted chains, as a function of~$L/\ell_p$. T
  $\moy[\text{tk}]\chicsq$ is below $\moy\chicsq$. The knot correction
  lowers the fluctuations. For $L/\ell_p=80$ the correction computed
  is $-7.5\%$. ($3\sigma$) shown for $\moy[\text{tk}]\chicsq$, error
  for~$\moy\chicsq$ small than symbol size.}
\label{figknot}
\end{center}
\end{figure}
In order to select the knotted chains, we have computed the Jones
polynomial~$V$ \cite{jones,kauff2}.  The choice of the Jones
polynomial is justified by the Jones conjecture, that is $V_K=1$ if
and only if~$K$ is the trivial knot.  We have used the algorithm of
Kauffman \cite{kauffman,kauff2} to calculate $V_K$ and perform the
classification of chains in function of their knots.  Recently it has
been noted that probability of knot formation is small \cite{grosberg}
in short chains.  Our simulations confirm this point~: for chains of 8
persistence lengths the proportion of knotted chain is around
$(5\pm1)\times10^{-4}$ when $f=0$.

To interpret a micromanipulation experiment one should perform
averages over an ensemble of chains with the same given knot.  In
practice one hopes the experiment is performed with the trivial knot.
The probability distribution of the writhe angle~$\chic$ has to be
modified, because we do not count the knotted configurations.  We have
computed~$\moy[\text{tk}]{\chicsq}$, where the label~``tk'' stands for
``trivial knot'' and compared it to~$\moy\chicsq$, for different
lengths with no force. Results are plotted on figure~\ref{figknot}.
The knot configurations that our algorithm generated lead to an
overestimated value for~$\moy\chicsq$.  The correction is about
$-7.5\%$ for the longest chains that we studied of~$L=80\ell_p$.  The
correction {\sl increases} the disagreement between calculations based
on Fuller's formulation of the writhe and the experimental curves.  It
is particularly instructive to compare figure \ref{figknot} and figure
\ref{modulofig} for the case $L/\ell_p =8$. Removing knots from the
ensemble of chains has a negligible effect on the writhe for such
short chains, however even for the most coarsely discretized chains
with $\ell_p/b =10$ the ratio
\begin{math}
\langle \chi_F ^2 \rangle /
\langle \chi_C ^2 \rangle 
\end{math} is larger than $2$.
We have also studied the evolution of this correction with the force.
The correction decreases rapidly with the force and is found to be
zero for forces $F\ell_p$ larger than a few~$k_BT$.



\section{Conclusions}
In this paper we have shown that the standard experimental geometry
does not torsionally confine a DNA strand so that under torque we
expect an series of equivalent low energy states separated by a
potential barrier. For the usual bead sizes and forces used in the
experiments this barrier is very high and the extended linking number
is conserved. This allows the use of an extended \ca-White formalism
in the calculation of the bead rotations.  We find that the Fuller and
\ca-White formulae{} give substantially different distribution
functions for the torsional fluctuations due to writhe. In contrast to
Bouchiat {\it et al.} we find that the topologically confined DNA
chain does not need an intermediate scale cut off to render the
response functions finite. The mathematical problems as to {\sl
  existence} of the torsional response functions occur at long
wavelengths; imposing a short wavelength cut off is the wrong solution
to this problem.

In experiments we expect several distinct regimes when working with
beads of size $d \gg \ell_p$. For very low forces, $F<k_BT/d$
torsional fluctuations are unbounded and it is not possible to define
the torsion-force-extension characteristics. In the regime
$k_BT/d<F<k_BT/\ell_p$ torsional fluctuations are bounded but must be
calculated using the full double integral representation of the
writhe. The Fuller formulation, even with an additional cut off,
substantially overestimates the writhe contribution to the torsional
response.  When $F\ell_p \sim k_BT$ a simple theory based on a Monge
representation expanded to quadratic order is unable to fit the data
due to strong corrections to scaling; here an expression based on the
Fuller formula can be expected to give a better description of the
response.  Finally, for very large forces, $F\gg k_BT/\ell_p$, the
\ca-White and Fuller formulae{} give the same result, however other
effects which are neglected here become important; a full theory
must treat the chiral nature of DNA and force induced denaturation.

\bibliography{vincent} 
\begin{thebibliography}{30}
\expandafter\ifx\csname natexlab\endcsname\relax\def\natexlab#1{#1}\fi
\expandafter\ifx\csname bibnamefont\endcsname\relax
  \def\bibnamefont#1{#1}\fi
\expandafter\ifx\csname bibfnamefont\endcsname\relax
  \def\bibfnamefont#1{#1}\fi
\expandafter\ifx\csname citenamefont\endcsname\relax
  \def\citenamefont#1{#1}\fi
\expandafter\ifx\csname url\endcsname\relax
  \def\url#1{\texttt{#1}}\fi
\expandafter\ifx\csname urlprefix\endcsname\relax\def\urlprefix{URL }\fi
\providecommand{\bibinfo}[2]{#2}
\providecommand{\eprint}[2][]{\url{#2}}

\bibitem[{\citenamefont{Strick et~al.}(1998)\citenamefont{Strick, Allemand,
  Bensimon, and Croquette}}]{ABCS}
\bibinfo{author}{\bibfnamefont{T.}~\bibnamefont{Strick}},
  \bibinfo{author}{\bibfnamefont{J.-F.} \bibnamefont{Allemand}},
  \bibinfo{author}{\bibfnamefont{D.}~\bibnamefont{Bensimon}}, \bibnamefont{and}
  \bibinfo{author}{\bibfnamefont{V.}~\bibnamefont{Croquette}},
  \bibinfo{journal}{Biophys. J.} \textbf{\bibinfo{volume}{74}},
  \bibinfo{pages}{2016} (\bibinfo{year}{1998}).

\bibitem[{\citenamefont{Smith et~al.}(1992)\citenamefont{Smith, Finzi, and
  Bustamante}}]{smith}
\bibinfo{author}{\bibfnamefont{S.~B.} \bibnamefont{Smith}},
  \bibinfo{author}{\bibfnamefont{L.}~\bibnamefont{Finzi}}, \bibnamefont{and}
  \bibinfo{author}{\bibfnamefont{C.}~\bibnamefont{Bustamante}},
  \bibinfo{journal}{Science} \textbf{\bibinfo{volume}{258}},
  \bibinfo{pages}{1122} (\bibinfo{year}{1992}).

\bibitem[{\citenamefont{Cluzel et~al.}(1996)\citenamefont{Cluzel, Lebrun,
  Heller, Lavery, Viovy, Chatenay, and Caron}}]{chatenay}
\bibinfo{author}{\bibfnamefont{P.}~\bibnamefont{Cluzel}},
  \bibinfo{author}{\bibfnamefont{A.}~\bibnamefont{Lebrun}},
  \bibinfo{author}{\bibfnamefont{C.}~\bibnamefont{Heller}},
  \bibinfo{author}{\bibfnamefont{R.}~\bibnamefont{Lavery}},
  \bibinfo{author}{\bibfnamefont{J.-L.} \bibnamefont{Viovy}},
  \bibinfo{author}{\bibfnamefont{D.}~\bibnamefont{Chatenay}}, \bibnamefont{and}
  \bibinfo{author}{\bibfnamefont{F.}~\bibnamefont{Caron}},
  \bibinfo{journal}{Science} \textbf{\bibinfo{volume}{271}},
  \bibinfo{pages}{792} (\bibinfo{year}{1996}).

\bibitem[{\citenamefont{Marko and Siggia}(1995)}]{markosiggia}
\bibinfo{author}{\bibfnamefont{J.}~\bibnamefont{Marko}} \bibnamefont{and}
  \bibinfo{author}{\bibfnamefont{E.~D.} \bibnamefont{Siggia}},
  \bibinfo{journal}{Phys. Rev. E} \textbf{\bibinfo{volume}{52}},
  \bibinfo{pages}{2912} (\bibinfo{year}{1995}).

\bibitem[{\citenamefont{Marko}(1997)}]{marko1}
\bibinfo{author}{\bibfnamefont{J.}~\bibnamefont{Marko}},
  \bibinfo{journal}{Phys. Rev.} \textbf{\bibinfo{volume}{E 55}},
  \bibinfo{pages}{1758} (\bibinfo{year}{1997}).

\bibitem[{\citenamefont{Moroz and Nelson}(1998)}]{nelson}
\bibinfo{author}{\bibfnamefont{J.}~\bibnamefont{Moroz}} \bibnamefont{and}
  \bibinfo{author}{\bibfnamefont{P.}~\bibnamefont{Nelson}},
  \bibinfo{journal}{Macromol.} \textbf{\bibinfo{volume}{31}},
  \bibinfo{pages}{6333} (\bibinfo{year}{1998}).

\bibitem[{\citenamefont{Vologodskii and Marko}(1997)}]{marko}
\bibinfo{author}{\bibfnamefont{A.}~\bibnamefont{Vologodskii}} \bibnamefont{and}
  \bibinfo{author}{\bibfnamefont{J.}~\bibnamefont{Marko}},
  \bibinfo{journal}{Biophys. J.} \textbf{\bibinfo{volume}{73}},
  \bibinfo{pages}{123} (\bibinfo{year}{1997}).

\bibitem[{\citenamefont{M\'ezard and Bouchiat}(1998)}]{MezardPRL}
\bibinfo{author}{\bibfnamefont{M.}~\bibnamefont{M\'ezard}} \bibnamefont{and}
  \bibinfo{author}{\bibfnamefont{C.}~\bibnamefont{Bouchiat}},
  \bibinfo{journal}{Phys. Rev. Lett.} \textbf{\bibinfo{volume}{80}},
  \bibinfo{pages}{1556} (\bibinfo{year}{1998}).

\bibitem[{\citenamefont{C\u{a}lug\u{a}reanu}(1959)}]{cal}
\bibinfo{author}{\bibfnamefont{G.}~\bibnamefont{C\u{a}lug\u{a}reanu}},
  \bibinfo{journal}{Rev. Math. Pures Appl.} \textbf{\bibinfo{volume}{4}},
  \bibinfo{pages}{5} (\bibinfo{year}{1959}).

\bibitem[{\citenamefont{White}(1969)}]{white}
\bibinfo{author}{\bibfnamefont{J.}~\bibnamefont{White}}, \bibinfo{journal}{Am.
  J. Math.} \textbf{\bibinfo{volume}{91}}, \bibinfo{pages}{693}
  (\bibinfo{year}{1969}).

\bibitem[{\citenamefont{Rossetto and Maggs}(2002)}]{us}
\bibinfo{author}{\bibfnamefont{V.}~\bibnamefont{Rossetto}} \bibnamefont{and}
  \bibinfo{author}{\bibfnamefont{A.~C.} \bibnamefont{Maggs}},
  \bibinfo{journal}{Phys. Rev. Lett} \textbf{\bibinfo{volume}{88}},
  \bibinfo{pages}{089801} (\bibinfo{year}{2002}).

\bibitem[{\citenamefont{Fuller}(1971)}]{fuller}
\bibinfo{author}{\bibfnamefont{F.~B.} \bibnamefont{Fuller}},
  \bibinfo{journal}{Proc. Nat. Acad. Sci. U.S.A.}
  \textbf{\bibinfo{volume}{68}}, \bibinfo{pages}{815} (\bibinfo{year}{1971}).

\bibitem[{\citenamefont{Fuller}(1978)}]{fuller2}
\bibinfo{author}{\bibfnamefont{F.~B.} \bibnamefont{Fuller}},
  \bibinfo{journal}{Proc. Nat. Acad. Sci. U.S.A.}
  \textbf{\bibinfo{volume}{75}}, \bibinfo{pages}{3557} (\bibinfo{year}{1978}).

\bibitem[{\citenamefont{Maggs}(2001)}]{acm}
\bibinfo{author}{\bibfnamefont{A.~C.} \bibnamefont{Maggs}},
  \bibinfo{journal}{J. Chem. Phys.} \textbf{\bibinfo{volume}{114}},
  \bibinfo{pages}{5888} (\bibinfo{year}{2001}).

\bibitem[{\citenamefont{Berry}(1987)}]{berry}
\bibinfo{author}{\bibfnamefont{M.}~\bibnamefont{Berry}},
  \bibinfo{journal}{Nature} \textbf{\bibinfo{volume}{326}},
  \bibinfo{pages}{277} (\bibinfo{year}{1987}).

\bibitem[{\citenamefont{Arnold and Keshin}(1991)}]{arnold}
\bibinfo{author}{\bibfnamefont{V.~I.} \bibnamefont{Arnold}} \bibnamefont{and}
  \bibinfo{author}{\bibfnamefont{B.~A.} \bibnamefont{Keshin}},
  \emph{\bibinfo{title}{Topological methods in hydrodynamics}}
  (\bibinfo{publisher}{Springer--Verlag}, \bibinfo{year}{1991}).

\bibitem[{\citenamefont{Feynman}(1987)}]{feynman}
\bibinfo{author}{\bibfnamefont{R.~P.} \bibnamefont{Feynman}},
  \emph{\bibinfo{title}{Elementary particles and the laws of physics: the 1986
  Dirac memorial lectures.}} (\bibinfo{publisher}{Cambridge University Press},
  \bibinfo{year}{1987}).

\bibitem[{\citenamefont{Antoine et~al.}(1991)\citenamefont{Antoine, Comtet,
  Desbois, and Ouvry}}]{orsay}
\bibinfo{author}{\bibfnamefont{M.}~\bibnamefont{Antoine}},
  \bibinfo{author}{\bibfnamefont{A.}~\bibnamefont{Comtet}},
  \bibinfo{author}{\bibfnamefont{J.}~\bibnamefont{Desbois}}, \bibnamefont{and}
  \bibinfo{author}{\bibfnamefont{S.}~\bibnamefont{Ouvry}}, \bibinfo{journal}{J.
  Phys. A.} \textbf{\bibinfo{volume}{24}} (\bibinfo{year}{1991}).

\bibitem[{\citenamefont{Klenin and Langowski}(2000)}]{linkwrithe}
\bibinfo{author}{\bibfnamefont{K.}~\bibnamefont{Klenin}} \bibnamefont{and}
  \bibinfo{author}{\bibfnamefont{J.}~\bibnamefont{Langowski}},
  \bibinfo{journal}{Biopolymers} \textbf{\bibinfo{volume}{54}},
  \bibinfo{pages}{307} (\bibinfo{year}{2000}).

\bibitem[{\citenamefont{Grosberg}(2000)}]{grosberg}
\bibinfo{author}{\bibfnamefont{A.~Y.} \bibnamefont{Grosberg}},
  \bibinfo{journal}{Phys. Rev. Lett.} \textbf{\bibinfo{volume}{85}},
  \bibinfo{pages}{3858} (\bibinfo{year}{2000}).

\bibitem[{\citenamefont{Doi and Edwards}(1992)}]{doiEdwards}
\bibinfo{author}{\bibfnamefont{M.}~\bibnamefont{Doi}} \bibnamefont{and}
  \bibinfo{author}{\bibfnamefont{S.}~\bibnamefont{Edwards}},
  \emph{\bibinfo{title}{The Theory of Polymer Dynamics}}
  (\bibinfo{publisher}{Clarendon, Oxford}, \bibinfo{year}{1992}).

\bibitem[{\citenamefont{Velikson et~al.}(1992)\citenamefont{Velikson, Garel,
  Niel, Orland, and Smith}}]{orland}
\bibinfo{author}{\bibfnamefont{B.}~\bibnamefont{Velikson}},
  \bibinfo{author}{\bibfnamefont{T.}~\bibnamefont{Garel}},
  \bibinfo{author}{\bibfnamefont{J.-C.} \bibnamefont{Niel}},
  \bibinfo{author}{\bibfnamefont{H.}~\bibnamefont{Orland}}, \bibnamefont{and}
  \bibinfo{author}{\bibfnamefont{J.~C.} \bibnamefont{Smith}},
  \bibinfo{journal}{J. Comput. Chem.} \textbf{\bibinfo{volume}{13}},
  \bibinfo{pages}{1216} (\bibinfo{year}{1992}).

\bibitem[{\citenamefont{de~Gennes}(1979)}]{degennes}
\bibinfo{author}{\bibfnamefont{P.-G.} \bibnamefont{de~Gennes}},
  \emph{\bibinfo{title}{Scaling concepts in polymer physics}}
  (\bibinfo{publisher}{Cornell University Press}, \bibinfo{address}{New York},
  \bibinfo{year}{1979}).

\bibitem[{\citenamefont{Bouchiat and M\'ezard}(2002)}]{reply}
\bibinfo{author}{\bibfnamefont{C.}~\bibnamefont{Bouchiat}} \bibnamefont{and}
  \bibinfo{author}{\bibfnamefont{M.}~\bibnamefont{M\'ezard}},
  \bibinfo{journal}{Phys. Rev. Lett.} \textbf{\bibinfo{volume}{88}},
  \bibinfo{pages}{089802} (\bibinfo{year}{2002}).

\bibitem[{\citenamefont{Levy}(1948)}]{levy}
\bibinfo{author}{\bibfnamefont{P.}~\bibnamefont{Levy}},
  \emph{\bibinfo{title}{Processus Stochastiques et Mouvement Brownien}}
  (\bibinfo{publisher}{Editions Jacques Gabay}, \bibinfo{year}{1948}).

\bibitem[{\citenamefont{Vologodskii}(2001)}]{vologo}
\bibinfo{author}{\bibfnamefont{A.~V.} \bibnamefont{Vologodskii}},
  \bibinfo{journal}{Molecular Biology} \textbf{\bibinfo{volume}{35}},
  \bibinfo{pages}{240} (\bibinfo{year}{2001}).

\bibitem[{\citenamefont{Klenin et~al.}(1989)\citenamefont{Klenin, Vologodskii,
  Anshelevich, Klisko, Dykhne, and Frank-Kamenetskii}}]{klenin}
\bibinfo{author}{\bibfnamefont{K.~V.} \bibnamefont{Klenin}},
  \bibinfo{author}{\bibfnamefont{A.~V.} \bibnamefont{Vologodskii}},
  \bibinfo{author}{\bibfnamefont{V.~V.} \bibnamefont{Anshelevich}},
  \bibinfo{author}{\bibfnamefont{V.~Y.} \bibnamefont{Klisko}},
  \bibinfo{author}{\bibfnamefont{A.~M.} \bibnamefont{Dykhne}},
  \bibnamefont{and} \bibinfo{author}{\bibfnamefont{M.~D.}
  \bibnamefont{Frank-Kamenetskii}}, \bibinfo{journal}{J. Biomol. Struct. Dyn.}
  \textbf{\bibinfo{volume}{6}}, \bibinfo{pages}{707} (\bibinfo{year}{1989}).

\bibitem[{\citenamefont{Jones}(1987)}]{jones}
\bibinfo{author}{\bibfnamefont{V.~F.~R.} \bibnamefont{Jones}},
  \bibinfo{journal}{Ann. of Math.} \textbf{\bibinfo{volume}{126}},
  \bibinfo{pages}{335} (\bibinfo{year}{1987}).

\bibitem[{\citenamefont{Kauffman}(1991)}]{kauff2}
\bibinfo{author}{\bibfnamefont{L.~H.} \bibnamefont{Kauffman}},
  \emph{\bibinfo{title}{Knots and physics}} (\bibinfo{publisher}{World
  Scientific}, \bibinfo{year}{1991}).

\bibitem[{\citenamefont{Kauffman}(1987)}]{kauffman}
\bibinfo{author}{\bibfnamefont{L.~H.} \bibnamefont{Kauffman}},
  \bibinfo{journal}{Topology} \textbf{\bibinfo{volume}{26}},
  \bibinfo{pages}{395} (\bibinfo{year}{1987}).

\end{thebibliography}
\end{document}